\documentclass[11pt]{article}
\newcounter{mnotecount}[section]

\renewcommand{\themnotecount}{\thesection.\arabic{mnotecount}}

\newcommand{\mnote}[1]
{\protect{\stepcounter{mnotecount}}$^{\mbox{\footnotesize
$
\bullet$\themnotecount}}$ \marginpar{
\raggedright\tiny\em
$\!\!\!\!\!\!\,\bullet$\themnotecount: #1} }

\usepackage{a4}
\usepackage{amssymb,latexsym}
\usepackage[mathscr]{eucal}
\usepackage{url}
\usepackage{ntheorem}

\newtheorem{Theorem} {\sc  Theorem\rm} [section]
\newtheorem{Corollary} {\sc  Corollary\rm} [section]
\newcommand{\tg}{\tilde{g}}
\newcommand{\ta}{\tilde{a}}
\newcommand{\tb}{\tilde{b}}
\newcommand{\tk}{\tilde{k}}%
\newcommand{\tnabla}{\tilde{\nabla}}%
\newcommand{\tp}{\tilde{p}}%
\newcommand{\ttau}{\tilde{\tau}}%
\newcommand{\tf}{\tilde{f}}%
\newcommand{\tM}{\widetilde{M}}

\newcommand{\tU}{\widetilde{U}}
\newcommand{\tR}{\widetilde{R}}
\newcommand{\tC}{\widetilde{C}}
\newcommand{\tG}{\widetilde{G}}
\newcommand{\tomega}{\tilde{\omega}}
\newcommand{\tT}{\widetilde{T}}

\newcommand{\bea}{\begin{eqnarray}}

\newcommand{\eea}{\end{eqnarray}}
\newcommand{\bean}{\begin{eqnarray*}}

\newcommand{\eean}{\end{eqnarray*}}
\newcommand{\be}{\begin{equation}}

\newcommand{\ee}{\end{equation}}

\begin{document}
\title{Isotropic cosmological singularities in spatially-homogeneous models with a cosmological constant}

\author{Paul Tod\thanks{{ E--mail}: \protect\url{paul.tod@st-johns.oxford.ac.uk}}
\\
Mathematical Institute and St John's College\\ Oxford}

\maketitle

\begin{abstract}
We prove well-posedness of the initial value problem for the Einstein equations for spatially-homogeneous cosmologies with data at an isotropic cosmological singularity, for which the matter content is either a cosmological constant with collisionless particles of a single mass (possibly zero) or a cosmological constant with a perfect fluid having the radiation equation of state. In both cases, with a positive cosmological constant, these solutions, except possibly for Bianchi-type-IX, will expand forever, and be geodesically-complete into the future.

\medskip

\end{abstract}

\section{Introduction}
\label{intro}
An isotropic cosmological singularity is one which may be removed by conformally rescaling the metric. More accurately, the physical space-time $(\tM,\tg)$ is conformally related to an unphysical space-time $(M,g)$ by $\tg=\Omega^2g$ where $\Omega$ is a function positive on $\tM$ and vanishing on a smooth space-like hypersurface $\Sigma$ in $M$, which provides a boundary to $\tM$ at which $g$ is regular but $\tg$ is singular.

In several cases of interest, it is possible to formulate a well-posed initial value problem in $M$ for the Einstein equations in $\tM$ by giving data at $\Sigma$. This was done for a perfect-fluid source with a linear equation of state in \cite{KA1}, for massless Einstein-Vlasov with spatial homogeneity in \cite{KA2} and for massless Einstein-Vlasov without the assumption of spatial homogeneity in \cite{KA3}. Steps towards the same end for a range of other matter models were taken in \cite{T1} and our purpose here is to carry through the proof of well-posedness in two cases with a cosmological constant $\Lambda$, namely massive or massless Einstein-Vlasov, and perfect fluid with the radiation equation of state. 

One motivation for this calculation is the recent appearance of long-time existence results for massive Einstein-Vlasov and a positive $\Lambda$ \cite{Lee}, \cite{TR}. In \cite{Lee}, it was shown that spatially-homogeneous cosmologies of Bianchi-types other than type-IX, once expanding, would expand forever and would be geodesically-complete into the future. (Type-IX \emph{may} do this but could recollapse instead.) In \cite{TR}, the corresponding result was proved for plane or hyperbolic symmetry. Putting what we find here together with Lee's result in \cite{Lee}, we may construct solutions of the Einstein equations with this matter model or with massless Einstein-Vlasov which expand away forever from an isotropic cosmological singularity to a future infinity or Scri (in the sense that the metric can be rescaled to add a future boundary at infinity; strictly speaking, to call this Scri, we should check for appropriate differentiability of the rescaled metric but, for the Einstein-Vlasov cases, we don't do that here). Given that result, it is natural to look for perfect fluid cosmologies with positive $\Lambda$ which also expand away forever from an isotropic cosmological singularity towards a Scri (this time, appropriately differentiable), and that is the other concern in this article. Both classes of models then provide fairly explicit examples of the cosmologies considered recently by Penrose \cite{P}: the conformal metric can be extended both through the initial singularity (or bang) and through the future infinity (or Scri). However, note that Penrose wanted more than this, namely that the conformal geometry at bang and Scri were essentially equivalent, and these models don't seem to have that property, though this isn't completely clear.

For the Einstein-Vlasov equations, our technique is simply to re-examine the proof of \cite{KA2}, with massive Vlasov in place of massless, and including a cosmological constant. As one would expect heuristically, and as was suggested without complete proof in \cite{T1}, these modifications are unimportant near the singularity. Our result is Theorem~\ref{thm1} of section~\ref{set}, that there is a well-posed initial value problem with data at the singularity, and the data is just the initial distribution function, subject to the vanishing dipole condition as in \cite{KA2}.

As Corollary~\ref{thm2}, we prove local-in-time existence for massless Einstein-Vlasov with $\Lambda$ near an isotropic singularity, and then for completeness prove long-time existence for this matter model for positive $\Lambda$ (Theorem~\ref{thm3}).

For the perfect fluid with radiation equation of state and $\Lambda$, we follow the calculation of \cite{KA1} to obtain local existence in time for a variety of spatially-homogeneous cosmologies, and then directly prove long-time existence.

\medskip

The plan of the paper is as follows. Section 2 is concerned with the Einstein-Vlasov results and Section 3 with the radiation results. 

We begin in sub-section 2.1 by collecting together some results on conformal rescaling for the massive Einstein-Vlasov equations with cosmological constant. In subsection 2.2, we specialise to the spatially-homogeneous case and write the Einstein equations as a first-order system of ordinary differential equations, coupled to the Vlasov equation which is still a partial differential equation. We discuss the splitting of the Einstein equations into constraints and evolution, and the result of imposing the constraints at the singularity. In subsection 2.3, we state the existence theorem, and show how it can be proved by the same techniques as in \cite{KA2} for the massless Einstein-Vlasov system without $\Lambda$. The proof can then be seen to carry through for massless Einstein-Vlasov with $\Lambda$. In subsection 2.4, closely following \cite{Lee}, we prove long-time existence for massless Einstein-Vlasov with positive $\Lambda$, so that for both the massive and massless case there exist solutions of the kind imagined by Penrose, with the conformal metric extendible through both the bang and Scri. However, it seems that even in the conformal geometry, the bang and Scri can be distinguished.

In Section 3, for the radiation results, we begin in subsection 3.1 with the Friedman-Robertson-Walker (FRW) metric. In terms of conformal time, the Einstein equations reduce to the Friedman equation for the scale factor. The scale factor grows from a simple zero at the bang to a simple pole at Scri, after a finite conformal time. For positive $\Lambda$, the Friedman equation has a remarkable symmetry in that the inverse of the scale factor satisfies the same equation as the scale factor, but with the density parameter and cosmological constant interchanged. This transformation also interchanges bang and Scri but, as we shall see, its existence depends on the absence of conformal curvature. In subsection 3.2, we consider the Bianchi type-III and Kantowski-Sachs metrics. We show that there are solutions expanding from an isotropic singularity at the bang. For positive $\Lambda$, the Kantowski-Sachs metrics may recollapse but the Bianchi-III metrics cannot and must expand to Scri, again after a finite conformal time. Finally, and most generally, in subsection 3.3 we consider metrics of Class A in the Bianchi classification. All types admit solutions expanding from an isotropic singularity at the bang. For positive $\Lambda$, all but type-IX, which may recollapse, expand to Scri, again after a finite conformal-time. The symmetry seen in FRW does not persist in the other examples: both the bang and Scri are umbilic in the conformal metric but the conformal metric is even in conformal-time about the bang but apparently not about Scri.

\section{The Einstein-Vlasov equations}
\subsection{Conformally rescaling the massive Einstein-Vlasov equations}
\label{sCR}
The physical space-time will be $(\tM,\tg)$, while the rescaled, unphysical space-time will be $(M,g)$ with
\be
\tg_{ab}=\Omega^2g_{ab}.\label{g1}
\ee
The physical space-time has an isotropic cosmological singularity if $\Omega$ vanishes at a smooth, space-like hypersurface $\Sigma$ in the unphysical space-time $M$. 

In kinetic theory, the matter content of space-time is
taken to consist of a
collection of particles which move on geodesics between
collisions.
States having a given mass~$m> 0$~define the seven-dimensional
submanifold $P_{m}$ of the one-particle phase-space $P=T^*\tM$. On ~$P_{m}$~we take~$(x^{a},p_{i})$~as local
coordinates, ~$p_{0}$~being determined by the
equation~$\tg^{ab}p_{a}p_{b}=m^{2}$, and the requirement that~$p_{a}$~be
future directed (conventionally for us, $a,b,c,..=0-3$ while
$i,j,k,..=1-3$). It is important for conformal rescaling to retain the notational convention $\tp^a=\tg^{ab}p_b$. 

The free-fall trajectories define on~$P$~a congruence
of curves, along which $H~=~\frac{1}{2}\tg^{ab}p_{a}p_{b}$~is constant. The geodesic spray~$\mathcal{L}_{\tg}$~is the vector
field tangent to these curves defined, in local coordinates, by
\begin{equation}
\mathcal{L}_{\tg}=\tg^{ab}p_{a}\frac{\partial}{\partial
x^{b}}-\frac{1}{2}p_{a}p_{b}\frac{\partial \tg^{ab}}{\partial
x^{c}}\frac{\partial}{\partial p_{c}}.
\label{l1}
\end{equation}

The cotangent space to~$\tM$ at a point $x$ is a flat Lorentzian manifold, and on
the submanifold $P_{m}(x)$ there exists an invariant volume measure $\tomega_{m}$ given by
\begin{equation}
\tomega_{m}=\frac{1}{\tp^{0}\sqrt{-\tg}}d^{3}p_{i}\label{f1}.
\end{equation}
In this article we shall be concerned with particles of a single mass, so that we shall henceforth omit the subscript $m$ on $\tomega$.
The distribution of particles and momenta is described by a non-negative scalar
function~$f=f(x^{a},p_{i})$~on~$P_m$, and the condition
that~$f$~represents a collisionless gas is simply
\begin{equation}
\mathcal{L}_{\tg}(f)=0,\label{V1}
\end{equation}
which is conveniently called the Vlasov equation. Note that~$f$~satisfies the
Vlasov equation if and only if it is constant along geodesics of $\tg$.\\

The stress-energy-momentum tensor due to these particles of mass~$m$~is
given by
\begin{equation}
\tT_{ab}(x)=\int_{P_{m}(x)}fp_{a}p_{b}\;\tomega,
\label{T1}
\end{equation}
and if the Vlasov equation is satisfied then
\begin{equation}
\tilde{\nabla}^{a}\tT_{ab}=0,\label{T2}
\end{equation}
where $\tilde{\nabla}$ is the metric covariant derivative for $\tg$.

The coupled Einstein-Vlasov equations, for the
metric $\tg_{ab}$ and the particle distribution function $f$ are therefore
\bean
\tG_{ab}&=&8\pi\int_{P_{m}}fp_{a}p_{b}\;\tomega\\
\mathcal{L}_{\tg}(f)&=&0
\eean
taking $G=c=1$. 

We shall investigate the transformation properties of the Einstein-Vlasov system
under conformal rescaling.
Under rescaling we choose, as our notation already implies, 
\[\tilde{p}_a=p_a,\]
so that the canonical one-form is unchanged:
\[\theta=\tilde{p}_adx^a=p_adx^a.\]
This then necessitates
\be
g^{ab}p_ap_b=\Omega^2\tg^{ab}p_ap_b=\Omega^2m^2,\label{l3}\ee
and at the approach to the initial surface $\Sigma$ in $M$ (which is the singularity in $\tM$), where $\Omega$ vanishes, the particles are effectively massless. Next
\[
\mathcal{L}_{\tg}f=\Omega^{-2}(\mathcal{L}_gf+m^2\Omega^2\Upsilon_a\frac{\partial f}{\partial p_a}),
\]
where, as usual, $\Upsilon_a=\Omega^{-1}\partial_a\Omega$, and the distribution function $f$ does {\emph{not}} change under rescaling. Thus, in $M$, the Vlasov equation from $\tM$ becomes the equation
\be
\mathcal{L}_gf+m^2\Omega^2\Upsilon_a\frac{\partial f}{\partial p_a}=0.\label{V2}\ee
Again, the mass term vanishes at $\Sigma$. For the volume-form on the mass-shell we obtain
\[\omega=\frac{1}{p^{0}\sqrt{-g}}d^{3}p_{i}=\Omega^2\tomega,\]
and we can define a rescaled energy-momentum tensor as
\[T_{ab}=\Omega^2\tT_{ab}=\int_{P_{m}}fp_{a}p_{b}\;\omega.\]
This will not be divergence-free, since in fact
\[\tg^{ab}\tnabla_a\tT_{bc}=\Omega^{-4}(g^{ab}\nabla_aT_{bc}-g^{ab}T_{ab}\Upsilon_c),\]
where $\nabla$ is the metric covariant derivative for $g$. Here the left-hand-side vanishes, which gives an equation for the divergence of $T_{ab}$ from the right-hand-side. Note the following equation for the trace: 
\[T=g^{ab}T_{ab}=\Omega^2m^2\int_{P_m} f\omega.\]
As one of the conditions for an isotropic cosmological singularity, we assume that $f$ extends to a smooth function on $T^*\Sigma$. Therefore the trace of the rescaled energy-momentum tensor vanishes at $\Sigma$, which is a reflection of the earlier observation that the mass of the particles is negligible there.

With the conformal transformation of the Einstein tensor taken
e.g. from \cite{KA1}, we can write down the conformal EV 
equations for~$g_{ab}$~and~$f$~as
\bea
\Omega^2G_{ab}&=&2\Omega\nabla_{a}\nabla_{b}\Omega-4\nabla_{a}\Omega\nabla_{b}\Omega-g_{ab}(2\Omega\square
\Omega-\nabla_{c}\Omega\nabla^{c}\Omega)\nonumber\\
&&+\frac{8\pi}{\sqrt{-g}}\int~fp_{a}p_{b}\frac{d^{3}p}{p^{0}}+\Lambda\Omega^4g_{ab}\label{G1}
\eea
together with (\ref{V2}).
%
\subsection{The spatially-homogeneous case}
\label{sSHC}
We briefly recall the formalism for spatially-homogeneous cosmologies (see e.g.\cite{hom}). These have a three-dimensional isometry group $G$ transitive on space-like hypersurfaces. Suppose $G$ has structure constants $C^i_{\;\;jk}$, for $i,j,k=1,2,3$, then one introduces a basis of left-invariant one-forms $\sigma^i$ satisfying
\be
d\sigma^i=\frac{1}{2}C^i_{\;\;jk}\sigma^j\wedge\sigma^k.
\label{m2}
\ee
The metric is given in this basis by the expression:
\begin{equation}
\tilde{g}_{ab}dx^adx^b=dt^2-\tilde{a}_{ij}(t)\sigma^{i}\sigma^{j}
\label{m1}
\end{equation}
where $t$ is proper-time on the
congruence orthogonal to the surfaces of homogeneity (which is necessarily geodesic) and $\tilde{a}_{ij}$ is the physical metric on the surfaces of homogeneity.

The Einstein-Vlasov equations reduce to a system of ordinary differential equations for $\tilde{a}_{ij}$, coupled with the Vlasov equation, which is a partial differential equation. We seek a well-posed formulation of these in the rescaled space-time. 
Following the massless case from \cite{KA2}, we introduce a new time-coordinate $\tau$ by $t=\tau^2$ and choose the conformal factor $\Omega=\tau$. (The justification for making these choices is that they lead to a well-posed initial value problem.) The rescaled metric is then
\bea
g_{ab}dx^adx^b&=&\Omega^{-2}\tg_{ab}dx^adx^b\nonumber\\
&=&4d\tau^2-a_{ij}(\tau)\sigma^i\sigma^j,
\label{m3}
\eea
where $a_{ij}=t^{-1}\tilde{a}_{ij}$, which we shall assume is regular at the initial surface $\Sigma$ where $\tau=0$. We shall systematically use $b^{ij}$ for the matrix inverse to $a_{ij}$.

We write the momentum in the basis of invariant one-forms as
\[p=2p_0d\tau+p_i\sigma^i,\]
and $p_a=(p_0,p_i)$. The normalisation (\ref{l3}) implies
\be
(p_0)^2=(p^0)^2=b^{ij}p_ip_j+m^2\tau^2.
\label{p3}
\ee
(Since $p^0=p_0$ with these conventions, we shall consistently use the form $p_0$ to avoid later confusion with initial values.) For the Vlasov equation, we first obtain the geodesic equation, which is:
\[\frac{dp_i}{d\lambda}=\gamma^a_{\;\;ib}p_a p^b,\]
where $\lambda$ is proper time for $g$, and the Ricci rotation coefficients $\gamma^a_{\;\;bc}$ are given by
\[\gamma^0_{\;\;ij}=0=\gamma^a_{\;\;b0}\]
and
\[\gamma^i_{jk}=\frac{1}{2}b^{im}(C^n_{\;\;jm}a_{nk}+C^n_{\;\;km}a_{nj}+C^n_{\;\;jk}a_{nm}).\]
The geodesic equation can therefore be written
\[\frac{dp_i}{d\lambda}=-C^k_{\;\;ij}b^{jm}p_kp_m.\]
The apparent sign-change here is because the spatial metric, with our conventions, is $-a_{ij}$. However, we shall henceforth observe the convention that indices $i,j,\ldots$ will be raised with $b^{ij}$ and lowered with $a_{ij}$.

The Vlasov equation simplifies in the spatially-homogeneous case because $\Omega$ is a function only of time, so that $\Upsilon$ has only a zero-component and the extra term drops out of (\ref{V2}). We assume that the distribution function will be homogeneous and distinguish the distribution function as a function of proper time $t$ from the distribution function as a function of conformal time $\tau$ by writing $\tf(t,p_i)=f(\tau,p_i)$. For the homogeneous distribution function the Vlasov equation reduces to the statement that $f$ is constant along geodesics of $g_{ab}$:
\[
\frac{df}{d\lambda}=\frac{1}{2}p_0\frac{\partial f}{\partial\tau}-\frac{\partial f}{\partial p_i}C^k_{\;\;ij}p_kp_lb^{lj}=0,
\]
or
\be
\frac{\partial f}{\partial\tau}=2(b^{mn}p_mp_n+m^2\tau^2)^{-1/2}C^k_{\;\;ij}p_kp_lb^{lj}\frac{\partial f}{\partial p_i}.
\label{V3}
\ee
Now we follow \cite{KA2} to find a first-order form of the equations, introducing a tensor $k_{ij}$ proportional to the second-fundamental form:
\bea
\frac{d}{d\tau}a_{ij}&=&k_{ij}\label{F1}\\
\frac{d}{d\tau}b^{ij}&=&-b^{im}b^{jn}k_{nm}\label{F2}\\
\frac{d}{d\tau}k_{ij}&=&-8R_{ij}+\frac{1}{\tau}(2Z_{ij}-2k_{ij}-b^{mn}k_{mn}a_{ij})+k_{im}k_{jn}b^{mn}\nonumber\\
&&-\frac{1}{2}(b^{mn}k_{mn})k_{ij}+8\Lambda\tau^2a_{ij}+\frac{32\pi}{\sqrt{a}}m^2a_{ij}\int \frac{fd^3p}{p_0}\label{F3}
\eea
where $R_{ij}$ is the spatial Ricci tensor, so that
\bean
8R_{ij}&=&-4C^k_{\;\;ck}(C^r_{\;\;tj}a_{ir}+C^r_{\;\;ti}a_{jr})b^{ct}\\
&&-4C^c_{\;\;ki}(C^k_{\;\;cj}+C^m_{\;\;tj}a_{cm}b^{kt})\\
&&+2C^m_{\;\;ks}C^r_{\;\;ct}a_{jm}a_{ir}b^{kt}b^{sc}.\eean
and
\be
Z_{ij}=\frac{1}{\tau}\left(\frac{32\pi}{\sqrt{a}}\int fp_ip_j\frac{d^3p}{p_0} -a_{ij}\right).\label{z1}\ee
We have introduced the new variables $Z_{ij}$, following \cite{KA2}, so that the pole in $\tau$ in (\ref{F3}) is no worse than first-order, but we therefore need an evolution equation for $Z_{ij}$. From the definition (\ref{z1}), using (\ref{V3}), we find
\bea
\frac{d}{d\tau}Z_{ij}&=&\frac{1}{\tau}[-Z_{ij}-k_{ij}-\frac{64\pi}{\sqrt{a}}C^k_{\;\;nr}b^{ln}\int\frac{\partial f}{\partial p_r}p_ip_jp_kp_l\frac{d^3p}{(p_0)^2}\label{F4}\\
&&-\frac{16\pi}{\sqrt{a}}\int((p_0)^2b^{mn}k_{mn}+b^{mq}b^{nr}k_{qr}p_mp_n+2m^2\tau) fp_ip_j\frac{d^3p}{(p_0)^3}],\nonumber
\eea
which again has no worse than a first-order pole. For later use, we write (\ref{F4}) in the form
\be
\frac{d}{d\tau}Z_{ij}=\frac{1}{\tau}(-Z_{ij}-k_{ij}-I_{ij}-J_{ij}^{\;\;mn}k_{mn})-L_{ij},
\label{F5}
\ee
where
\bea
I_{ij}&=& \frac{64\pi}{\sqrt{a}}C^k_{\;\;nr}b^{ln}\int\frac{\partial f}{\partial p_r}p_ip_jp_kp_l\frac{d^3p}{(p_0)^2},  \label{F6}\\
J_{ij}^{\;\;mn}&=& \frac{16\pi}{\sqrt{a}}\int f((p_0)^2b^{mn}
+b^{mq}b^{nr}p_qp_r)p_ip_j\frac{d^3p}{(p_0)^3},  \label{F7}\\
L_{ij}&=&\frac{32\pi m^2\tau}{\sqrt{a}}\int fp_ip_j\frac{d^3p}{(p_0)^3}. \label{F8}
\eea

From (\ref{F1}) and (\ref{F2}), 
\[\frac{d}{d\tau}(a_{ij}b^{jm}-\delta_i^m)=-k_{rn}b^{rm}(a_{ij}b^{jn}-\delta_i^n),\]
so that, by a Gronwall estimate, if $a_{ij}b^{jm}=\delta_i^m$ at $\tau=0$, then this holds for all $\tau$.

For the Hamiltonian constraint we calculate:
\be
C:=\frac{16\pi}{\sqrt{a}}\int fp_0d^3p-\frac{3}{2}-\tau^2(R-\frac{1}{16}k^{ij}k_{ij}+\frac{1}{16}k^2)-\frac{1}{2}k\tau+2\Lambda\tau^4=0,
\label{C1}
\ee
while for the momentum constraint we calculate:
\be
C_i:=\frac{32\pi}{\sqrt{a}}\int fp_id^3p-\tau^2(b^{mn}k_{nj}C^j_{\;\;mi}+b^{mn}k_{ni}C^j_{\;\;mj})=0.
\label{C2}
\ee
We calculate for the evolution of the constraints:
\bean
\partial_\tau(\tau^2aC)&=&C_ib^{ij}C^m_{\;\;jm},\\
\partial_\tau(\sqrt{a} C_i)&=&0.
\eean
so that the constraints are satisfied at all times if satisfied initially.

We use a superscript $0$ to denote the value of a quantity initially. To satisfy the constraints initially we need for (\ref{C2}) a condition on the initial distribution function $f^0(p_i)$:
\be
\int f^0p_id^3p=0,
\label{C3}
\ee
while from (\ref{C1})
\be
\int f^0p_0d^3p=\frac{3\sqrt{a^0}}{32\pi},
\label{C4}
\ee
which is essentially just a normalisation condition. From (\ref{z1}), the finiteness of $Z_{ij}$ at $\tau=0$ determines the initial metric, as in \cite{KA2}, through the equation
\be
a^0_{ij}=\frac{32\pi}{\sqrt{a^0}}\int f^0p_ip_j\frac{d^3p}{p_0}.
\label{C5}
\ee
(The proof that this determines $a^0_{ij}$ from $f^0$ goes through just as in the earlier reference.) Now (\ref{C4}) can be seen as just the trace of (\ref{C5}), since at $\tau=0$, $(b^0)^{ij}p_ip_j=(p_0)^2$.

For the initial values $k^0_{ij}$ and $Z^0_{ij}$ of $k_{ij}$ and $Z_{ij}$ we use (\ref{F3}) and (\ref{F5}) evaluated at $\tau=0$ to obtain the system:
\bean
2Z^0_{ij}-2k^0_{ij}-a^0_{ij}(b^0)^{mn}k^0_{mn}&=&0,\\
Z^0_{ij}+k^0_{ij}+(J^0)_{ij}^{\;\;mn}k^0_{mn}&=&-I^0_{ij}.
\eean
The calculation is precisely as in \cite{KA2} and we find $k^0_{ij}=Z^0_{ij}$ and an equation for $k^0_{ij}$:
\be
\chi^0_{\;ijlm}(b^0)^{lp}(b^0)^{lq}k^0_{pq}-2k^0_{ij}=\chi^0_{\;ij},
\label{C6}
\ee
where
\bean
\chi^0_{\;ijkl}&=&\frac{16\pi}{\sqrt{a^0}}\int f^0p_ip_jp_kp_l\frac{d^3p}{(p_0)^3}\\
\chi^0_{\;ij}&=&\frac{64\pi}{\sqrt{a^0}}\int f^0(C^k_{\;\;in}(b^0)^{nm}p_jp_kp_m+C^k_{\;\;jn}(b^0)^{nm}p_ip_kp_m\\
&&\qquad \qquad +C^k_{\;\;kn}(b^0)^{nm}p_ip_jp_m)\frac{d^3p}{(p_0)^2},
\eean
which can be thought of as related to the fourth and third moments respectively of the initial distribution function, much as (\ref{C5}) relates the initial metric to its second moment, and (\ref{C3}) is the vanishing of the first moment. These two equations are just as in \cite{KA2}, except with a minor typo corrected. As argued there, they give a unique $k^0_{ij}$ from $f^0$, which is also trace-free, so that $Z^0_{ij}=k^0_{ij}$.

\subsection{The existence theorem}
\label{set}
In this section, our purpose is to put the equations into a form in which the existence theorem of \cite{KA2}, which was in turn based on work in \cite{RS} and \cite{R1}, can be applied. We shall obtain short-time existence and well-posed-ness for the massive Einstein-Vlasov equations with cosmological constant and data at an isotropic singularity. The proof is easily simplified to cover existence for the limit of massless Einstein-Vlasov. The result is:
\begin{Theorem}\nonumber\label{thm1}
Let $G$ be a 3-dimensional Lie group of some Bianchi type, and let $(\sigma^i)$ be a basis of left-invariant one-forms on $G$; let $f^0(p_i)$ be a smooth function on the cotangent bundle of $G$, where $p_i$ are the components of $p$ in the chosen frame; suppose that $f^0$ is compactly supported, supported outside a neighbourhood of the origin, and satisfies the constraint (\ref{C3}); let $m$ be a positive constant; then there exists a positive real $T$ and exactly one smooth solution $(\tf(t,p_i), \tg_{ab})$ of the space-time Einstein-Vlasov equations on $G\times (0,T]$ with an isotropic singularity at $t=0$ and $\tf(t,p_i)\rightarrow f^0(p_i)$ as $t\rightarrow 0$.
\end{Theorem}
We start by transforming (\ref{F3}) and (\ref{F4}). The transformation makes repeated use of the following elementary formula:
\be
\frac{1}{\tau}(F(\tau)-F(0))=\int_0^1\dot{F}(s\tau)ds.\label{e2}\ee
Introduce new variables $\kappa_{ij}$ and $\zeta_{ij}$ by
\bean
\kappa_{ij}&=&k_{ij}-k^0_{ij}\\
\zeta_{ij}&=&Z_{ij}-Z^0_{ij}
\eean
so that $\kappa_{ij}$ and $\zeta_{ij}$ are initially zero, then (\ref{F3}) leads to
\bea
\frac{d}{d\tau}\kappa_{ij}&=&\frac{1}{\tau}(2\zeta_{ij}-2\kappa_{ij}-a^0_{ij}(b^0)^{mn}\kappa_{mn})\nonumber\\
&&+(k^0_{im}+\kappa_{im})(k^0_{jn}+\kappa_{jn})b^{mn}-\frac{1}{2}b^{mn}(k^0_{mn}+\kappa_{mn})(k^0_{ij}+\kappa_{ij})\nonumber\\
&&+P_{ij}^{\;\;mn}(k^0_{mn}+\kappa_{mn})-8R_{ij}+8\Lambda\tau^2a_{ij},\label{e1}
\eea
where, following (\ref{e2}), 
\[P_{ij}^{\;\;mn}=\int_0^1(b^{mn}(k^0_{ij}+\kappa_{ij})-a_{ij}b^{mp}b^{nq}(k^0_{pq}+\kappa_{pq}))(s\tau)ds.\]
Similarly (\ref{F5}) leads to
\be
\frac{d}{d\tau}\zeta_{ij}=\frac{1}{\tau}(-\zeta_{ij}-\kappa_{ij}-(J^0)_{ij}^{\;\;mn}\kappa_{mn})-L_{ij}+S^{(1)}_{ij}+S^{(2)}_{ij},
\label{e3}
\ee
where
\bean
S^{(1)}_{ij}&=&-\int_0^1\dot{I}_{ij}(s\tau)ds\\
S^{(2)}_{ij}&=&-\left(\int_0^1\dot{J}_{ij}^{\;\;pq}(s\tau)ds\right) (k^0_{pq}+\kappa_{pq}).
\eean
Note that $\dot{I}_{ij}$ and $\dot{J}_{ij}^{\;\;pq}$ can be calculated from (\ref{V3}), (\ref{F1}) and (\ref{F2}) in terms of $a$, $b$, $k$ and $f$. 

Equations (\ref{e1}) and (\ref{e3}) can be written
\be
\frac{du}{d\tau}+\frac{1}{\tau}Nu=G(a,b,u,f),
\label{e4}
\ee
where $u=(\kappa_{ij},\; \zeta_{ij})$ and $N$ is a constant matrix constructed from $a^0_{ij}$ and $(b^0)^{ij}$. In fact $N$ is exactly as in \cite{KA2} so that, as there, it is diagonalisable and all eigenvalues have positive real part. Suppose it is diagonalised by the change of coordinates $(y^\alpha)=y=Lu$, with a constant matrix $L$, then (\ref{e4}) becomes
\be
\frac{dy^\alpha}{d\tau}+\frac{1}{\tau}\lambda^\alpha y^\alpha=H^\alpha(a,b,y,f),\label{e5}
\ee
where the eigenvalues $\lambda^\alpha$ all have positive real part, and the summation convention is suspended. The solution of a system like (\ref{e5}) is dealt with in \cite{RS}. It can be solved in a form convenient for iteration by the explicit formula
\[y_\alpha=\tau^{-\lambda_\alpha}\int_0^\tau \sigma^{\lambda_\alpha}H^\alpha(a,b,y,f)d\sigma.\]
Equation (\ref{V3}) is solved by the method of characteristics, but first we use a device to deal with the origin in $p_i$, which is significant for bounding the integrals over $p_i$ in (\ref{F3}) and (\ref{F4}) (this is exactly as in \cite{KA2} which in turn follows \cite{R1}). By assumption, $f^0$ has compact support and is bounded away from the origin. Suppose that $f^0=0$ for $|p|<B_1$ and choose $B_2<B_1$ and $B_3<B_2$ and a smooth function $\phi(r)$ with $\phi=0$ for $r<B_3$, $\phi=1$ for $r>B_2$ and $0\leq\phi\leq 1$ elsewhere. Suppose $b^{ij}$ is known, and solve
\be
\frac{dP_i}{d\lambda}=-2\phi(|P|)(b^{mn}P_mP_n+m^2\tau^2)^{1/2}b^{jl}C^k_{\;\;ij}P_kP_l
\label{V4}
\ee
for $P_i(\lambda;\tau,p_j)$ subject to $$P_i(\tau;\tau,p_j)=p_i.$$Then (\ref{V3}), with a factor $\phi$ on the right-hand-side, is solved by 
$$f(\tau, p_i)= f^0(P_i(0;\tau,p_j)).$$ 
As long as $f$ has support where $\phi=1$ this is the same equation.

\medskip

The method of solution for the whole system, as in \cite{KA2}, is now an iteration. Given smooth iterates $(a_n, b_n, y_n, f_n)$:
\begin{itemize}
\item
solve (\ref{V4}) using $b_n$ and obtain $f_{n+1}$; this is smooth if $b_n$ and $f^0$ are;
\item
solve (\ref{F1}), (\ref{F2}) and (\ref{e5}) with $(a_n, b_n, y_n, f_{n+1})$ on the right-hand-side for smooth $(a_{n+1},b_{n+1},y_{n+1})$;
\item
following the argument in \cite{KA2} (equations (92)-(97) there), show inductively that the iterates are bounded uniformly in $n$ and likewise for the support of $f_n$, for some interval in $\tau$, say $[0,T)$; 
\item
now the proof of convergence follows the argument in \cite{KA2}, the only change being an extra term in equation (100) there which becomes 
\[|y_{n+1}-y_n|\leq C\int_0^\tau\{(|a_n-a_{n-1}|+|b_n-b_{n-1}|+|y_n-y_{n-1}|+\|f_{n+1}-f_{n}\|_{\infty})(s)\]
\[+\left(\int_0^1(|a_n-a_{n-1}|+|b_n-b_{n-1}|+|y_n-y_{n-1}|+\|f_{n+1}-f_{n}\|_{\infty})(sp)dp\right)\}ds;\]
\item
finally uniqueness goes through as before.


\end{itemize}

\medskip

These solutions exist for at least a finite conformal time, and therefore for a finite proper time, when they provide data for the existence theorem of Lee \cite{Lee}. Thus for Bianchi types other than type-IX, they will exist forever in proper time and have the asymptotic behaviour found by Lee. The metric can be rescaled to add the future boundary so, in particular, these solutions expand from a bang to a Scri and the conformal metric is extendible through both.

To obtain a similar result for the massless case, we first need local existence with data at the bang. This was done without $\Lambda$ in \cite{KA2}. With $\Lambda$, we seek to set $m=0$ in the above proof. Note the explicit appearances of $m$ in (\ref{V3}), (\ref{F3}) and (\ref{F4}) and recall its implicit appearance in $p_0$ via (\ref{p3}). The first three can harmlessly be set to zero (there is an effect on the rate at which the support of $f$ spreads, but this is covered by the calculation in \cite{KA2}). The fourth can be set to zero since the support of $f$ is bounded away from the origin, so that the integrals over $p$ are finite. Thus the same proof goes through for massless Einstein-Vlasov and we have:
\begin{Corollary}\label{thm2}
Theorem \ref{thm1} holds for $m=0$
\end{Corollary}
To find a Scri in the massless case, we next seek to modify the proof of long-time existence in \cite{Lee} to cover $m=0$.

\subsection{Long-time existence for massless Einstein-Vlasov with $\Lambda>0$} 
Long-time existence for some spatially-homogeneous cosmologies whose source is massive Einstein-Vlasov with positive $\Lambda$ was proved by Lee \cite{Lee} with the aid of a continuation result of Rendall \cite{R1}. We now show how this can be extended to massless Einstein-Vlasov with positive $\Lambda$. We revert to the space-time metric (\ref{m1}) and proper-time $t$ and by transforming the system (\ref{F1})-(\ref{F3}) and setting $m=0$, obtain the system
\bea
\frac{d}{dt}\ta_{ij}&=&\tk_{ij}\label{FF1}\\
\frac{d}{dt}\tb^{ij}&=&-\tb^{im}\tb^{jn}\tk_{nm}\label{FF2}\\
\frac{d}{dt}\tk_{ij}&=&-2R_{ij}+\tk_{im}\tk_{jn}\tb^{mn}-\frac{1}{2}(\tb^{mn}\tk_{mn})\tk_{ij}+2\Lambda\ta_{ij}\nonumber\\
&&+\frac{16\pi}{\sqrt{\ta}}\int \tf p_ip_j\frac{d^3p}{\tp^0}\label{FF3}
\eea
where $R_{ij}$ is as before and
\[\tp^0=(\tb^{mn}p_mp_n)^{1/2}.\]
The Vlasov equation becomes
\be
\frac{\partial \tf}{\partial t}=(\tb^{mn}p_mp_n)^{-1/2}C^k_{\;\;ij}p_kp_l\tb^{lj}\frac{\partial \tf}{\partial p_i}.
\label{V5}
\ee
These equations can be checked against the system in \cite{Lee} (with some minor changes of convention), and we want to see that the proof of long-time existence given there carries over.


The Hamiltonian constraint is the equation
\be
\tC:=\frac{16\pi}{\sqrt{\ta}}\int \tf\tp^0d^3p-\tb^{ij}R_{ij}+2\Lambda+\frac{1}{4}(\tk^{ij}\tk_{ij}-\tk^2)=0,
\label{FF4}
\ee
and the momentum constraint follows from (\ref{C2}) as
\be
\tC_i:=t^{-3/2}C_i=\frac{32\pi}{\sqrt{\ta}}\int \tf p_id^3p-2\tb^{mn}(C^j_{\;\;mi}\tk_{nj}+C^j_{\;\;mj}\tk_{ni})=0.
\label{FF7}
\ee
\begin{Theorem}\label{thm3}
Given data $(\ta_{ij},\tb^{ij},\tk_{ij},\tf)$ at some time $t_0$, with $\tk>0$ and satisfying the constraints (\ref{FF4}) and (\ref{FF7}), then for Bianchi types other than type-IX, there exists a unique solution of the Einstein equations with this data, which furthermore exists for all $t\geq t_0$. The solution admits a Scri.
\end{Theorem}

The proof is very similar to that in \cite{Lee}. 
From the trace of (\ref{FF3}), with the aid of (\ref{FF4}), we obtain
\be
\frac{d\tk}{dt}=2\Lambda-\frac{1}{2}\tk^{ij}\tk_{ij}-\frac{16\pi}{\sqrt{\ta}}\int \tf\tp^0d^3p.\label{FF6}\ee
Decompose $\tk_{ij}$ as
\[\tk_{ij}=\sigma_{ij}+\frac{1}{3}\tk\ta_{ij},\]
so that $\sigma_{ij}$ is trace-free, then (\ref{FF4}) can be rearranged to give
\be
\frac{16\pi}{\sqrt{\ta}}\int \tf\tp^0d^3p-\tb^{ij}R_{ij}+\frac{1}{4}\sigma^{ij}\sigma_{ij}=-2\Lambda+\frac{1}{6}\tk^2.
\label{FF5}
\ee
Following \cite{Lee}, we restrict to spatially-homogeneous metrics not of Bianchi type-IX, then these have $\tb^{ij}R_{ij}\leq 0$, so that the left-hand-side in (\ref{FF5}) is non-negative. Putting this with (\ref{FF6}) we obtain
\[\frac{d\tk}{dt}\leq 2\Lambda-\frac{1}{6}\tk^2\leq 0,\]
for all but type-IX. Now $\tk$ is initially positive and we can integrate this last equation to find
\be
6H\leq\tk\leq 6H\frac{(1+e^{-2H t})}{(1-e^{-2H t})},
\label{B1}
\ee
where $H=\sqrt{\Lambda/3}$, which is the $\gamma$ of \cite{Lee} . This bounds $\tk$ above and below for positive $t$ and forces $\tk\rightarrow 6H$ as $t\rightarrow\infty$. From (\ref{FF1})
\[\frac{d}{dt}\log\ta=\tk,\]
so that $\log\ta$ is also bounded above and below for positive $t$ and tends to a linear function of $t$. From (\ref{FF5}),
\[\frac{1}{4}\sigma_i^{\;\;j}\sigma_j^{\;\;i}\leq-2\Lambda+\frac{1}{6}\tk^2\leq \frac{8\Lambda e^{-2H t}}{(1-e^{-2H t})^2},\]
so that $\sigma_i^{\;\;j}$ is bounded and tends to zero, and we may follow the argument in \cite{Lee} to find the asymptotic form of the metric. The solutions exist for ever, are complete in the future and can be rescaled to add future infinity.

\medskip
 



Putting this with the Corollary \ref{thm2}, we obtain solutions of massless Einstein-Vlasov with positive $\Lambda$ which expand forever from an isotropic singularity. From the work of Lee, \cite{Lee}, it follows that the distribution function has a limit at Scri, but there seems no reason to expect that (\ref{C3}) of (\ref{C5}) will hold at Scri. If this is so, then for this matter model bang and Scri have different conformal properties.

\section{Radiation with $\Lambda$}
In this section, the object is to prove existence of examples of spatially-homogeneous cosmological models, whose source is a perfect fluid with the radiation equation of state and a cosmological constant, with data given at an isotropic singularity. We also show that, for positive $\Lambda$, some of these expand for ever (in proper time) and have a smooth Scri. We present a series of increasingly complex examples, starting with a simple case.

\subsection{The example of FRW}
The FRW space-time metric is
\[\tg=dt^2-R^2(t)d\sigma_k^2\]
where $R(t)$ is the scale factor and $d\sigma_k^2$ stands for the metric of a 3-dimensional space of constant curvature which is positive, negative or zero according as $k$ is $1$, $-1$ or $0$.

With a perfect fluid source and the radiation equation of state, $p=\rho/3$, the Einstein field equations reduce to the conservation equation, which integrates to give
\be
\rho R^4=m,
\label{R1}
\ee
where $m$ is a (positive) constant of integration, together with the Friedman equation
\be
\dot{R}^2+k=\frac{8\pi G}{3c^2}\rho R^2+\frac{\Lambda}{3}R^2.
\label{R2}
\ee
Here the over-dot indicates differentiation with respect to proper time $t$. We introduce conformal time $\tau$ by 
\be
d\tau=dt/R(t)
\label{R4}
\ee
and denote differentiation with respect to it by a prime. Then using (\ref{R1}) in (\ref{R2}) and setting $\frac{8\pi G}{c^2}=1$ we find
\be
(R')^2 = \frac{m}{3}-kR^2+\frac{\Lambda}{3}R^4.
\label{R3}
\ee
This equation has a unique solution with $R=0$, $R'>0$ at $\tau=0$, corresponding to an initial singularity. If we assume that either $k$ is negative or zero or that $k=1$ but $m\Lambda>9/4$ then this solution blows up in finite $\tau$-time, at say $\tau=\tau_F$ given by
\[\tau_F=\int_0^\infty \frac{dR}{(\frac{m}{3}-kR^2+\frac{\Lambda}{3}R^4)^{1/2}}.\]
In fact $R$ has a simple pole $R\sim(H(\tau_F-\tau))^{-1}$ where, as in the previous section, $H=\sqrt{\Lambda/3}$. Thus, by (\ref{R4}), in terms of proper time $R\sim H^{-1}e^{H t}$ and $\tau_F$ corresponds to future infinity or Scri.

Equation (\ref{R3}) also has a striking symmetry: if we replace $R$ by $\tR=R^{-1}$ we obtain the same equation but with $m$ and $\Lambda$ interchanged:
\[(\tR')^2 = \frac{\Lambda}{3}-k\tR^2+\frac{m}{3}\tR^4.\]
Evidently this symmetry interchanges Scri and the initial singularity, while also interchanging the constants $m$ and $\Lambda$. The conformal metric extends smoothly through both ends, but this is perhaps unsurprising as the metric is conformally flat.

In subsequent examples we find that the conformal extension property is preserved, but not the symmetry interchanging Scri and bang, which is disrupted by the presence of conformal curvature.

\subsection{Bianchi-III and Kantowski-Sachs models}
These have a space-time metric which can be parametrised as follows:
\be
\tg=dt^2-(RS^2)^2dz^2-(RS^{-1})^2(d\theta^2+f_k(\theta)^2d\phi^2).
\label{ks1}
\ee
Here $k=+1$ and $f_1=\sin\theta$ for the Kantowski-Sachs metric, and $k=-1$ and $f_{-1}=\sinh\theta$ for the Bianchi type-III metric.

This choice of parametrisation allows the introduction of conformal time $\tau$ just as in (\ref{R4}) and the solution of the conservation equation just as in (\ref{R1}). The conformally rescaled metric
\[g:=R^{-2}\tg=d\tau^2-S^4dz^2-S^{-2}(d\theta^2+f_k(\theta)^2d\phi^2)\]
is evidently smooth wherever $S$ is smooth and non-zero, and we shall see examples where this includes the initial singularity and Scri.

The Einstein equations can be written as the system
\bea
3\frac{S''}{S}+6\frac{R'S'}{RS}-3\left(\frac{S'}{S}\right)^2-kS^2&=&0\label{ks2}\\
3\frac{R''}{R}+3\left(\frac{S'}{S}\right)^2+kS^2-2\Lambda R^2&=&0\label{ks3}
\eea
where these have been simplified with the aid of the Hamiltonian constraint, which in turn can be written as
\be
3\left(\frac{R'}{R}\right)^2= 3\left(\frac{S'}{S}\right)^2-kS^2+\frac{m}{R^2}+\Lambda R^2.
\label{ks5}
\ee
We shall see below that these have solutions with initial data $(R,R',S,S')=(0,\sqrt{m/3},S_0,0)$ at (say) $\tau=0$. Suppose this is true, then given $\Lambda$, this is a 2-parameter family expanding from an isotropic cosmological singularity. The Kantowski-Sachs metric could recollapse but the Bianchi type-III metric cannot, as we see as follows: introduce
\be
Q=3\left(\frac{S'}{S}\right)^2-kS^2,
\label{ks8}
\ee
which is manifestly non-negative for $k=-1$. Then by (\ref{ks2})
\[Q'=-12\frac{R'}{R}\left(\frac{S'}{S}\right)^2\leq 0\]
so that, with the chosen initial conditions, 
\[0\leq Q\leq S_0^2.\]
Therefore $S$ and $\frac{S'}{S}$ are bounded for all time. Write (\ref{ks5}) as
\be
3\left(\frac{R'}{R}\right)^2=\frac{m}{R^2}+\Lambda R^2+Q.
\label{ks6}
\ee
The right-hand-side is strictly positive, so that $R'$, positive initially, never vanishes. Also $\frac{R'}{R}$ is bounded as long as $R$ is. Thus solutions exist until $R$ diverges. This happens after finite conformal time, as we see by comparing $R$ with the solution $L$ of the equation
\[3(L')^2=m+\Lambda L^4,\;\;L(0)=0,\;\;L'(0)>0.\]
Then $R\geq L$ but $L$ diverges in a time
\[\sqrt{3}\int_0^\infty\frac{dL}{(m+\Lambda L^4)^{1/2}},\]
which is finite. We shall return to the asymptotic form after proving existence
of solutions with data as claimed. For this, we put the system of Einstein equations into a first-order Fuchsian form, like (\ref{e4}). Set
\be
R=\tau e^U,\;\;S=e^\Sigma,
\label{ks7}
\ee
when (\ref{ks2}) and (\ref{ks3}) become
\bean
\Sigma'&=&Z\\
U'&=&W\\
Z'&=&-\frac{2}{\tau}Z-2WZ+\frac{k}{3}\;e^{2\Sigma}\\
W'&=&-\frac{2}{\tau}W-Z^2-W^2-\frac{k}{3}\;e^{2\Sigma}+\frac{2\Lambda}{3}\tau^2e^{2U}
\eean
while the Hamiltonian constraint becomes
\[m=e^{2U}\left(3+6\tau W+\tau^2(3W^2-3Z^2+ke^{2\Sigma}-\Lambda\tau^2e^{2U})\right),\]
which can be interpreted just as the statement that the right-hand-side is constant, which evaluation at $\tau=0$ shows to be positive. (In this setting, the momentum constraint is vacuously satisfied.)

This is the desired first-order Fuchsian form and satisfies the conditions for Theorem 1 of \cite{RS}, so that it has smooth solutions given data $(\Sigma, U,Z,W)=(\Sigma_0,U_0,0,0)$ at $\tau=0$. Clearly the two free parameters correspond to $m$ and $S_0$ as claimed.

It is straightforward to show inductively that as power series about $\tau=0$ $U$ and $\Sigma$ have only even powers of $\tau$ so that $S$ is even while $R$ is odd. In particular the initial singularity is umbilic in the conformal metric. We try the corresponding expansion at Scri: if $R$ diverges at say $\tau=\tau_F$ then introduce $\tR=R^{-1}$ so that (\ref{ks6}) becomes
\be
3(\tR')^2=\Lambda+Q\tR^2+m\tR^4,
\label{ks9}
\ee
and, as $\tau\rightarrow\tau_F$, $\tR\rightarrow 0$ and $\tR'\rightarrow-\sqrt{\Lambda/3}=-H$. Thus $R$ has a power series (strictly, a Laurent series) beginning with a first-order pole. We can write (\ref{ks2}) as
\be
\left(R^2\frac{S'}{S}\right)'=\frac{k}{3}R^2S^2=-\frac{1}{3}R^2S^2
\label{ks10}
\ee
in the Bianchi type-III case, so that
\[R^2\frac{S'}{S}=-\frac{1}{3}\int_0^\tau R^2(\sigma)S^2(\sigma)d\sigma,\]
from which, given the asymptotic behaviour of $R$ just found, $\frac{S'}{S}\rightarrow 0$ and $S\rightarrow S_{\infty}$ say, as $\tau\rightarrow\tau_F$. Now differentiate (\ref{ks9}) again to find
\[3\tR''=\left(Q+6\left(\frac{S'}{S}\right)^2\right)\tR+2m\tR^3,\]
so that $\tR''\rightarrow 0$ as $\tau\rightarrow\tau_F$, and differentiate again to find
\[\tR'''\rightarrow -\frac{1}{3}H\;S_{\infty}^2.\]
From the integral solution of (\ref{ks10}) we find $S''$ and $S'''$ finite at $\tau_F$ and, if we set $\ttau=\tau_F-\tau$, then power series begin
\bean
\tR&=&H(\ttau+\frac{S_{\infty}^2}{18}\;\ttau^3+\cdots)\\
R&=&H^{-1}\ttau^{-1}(1-\frac{S_{\infty}^2}{18}\;\ttau^2+\cdots)\\
S&=&S_\infty+\frac{S_{\infty}^3}{18}\;\ttau^2+S_3\;\ttau^3+\cdots.
\eean
Here $S_{\infty}$ and $S_3$ are fixed by initial conditions but not in a way we can find explicitly. From the point of view of Scri, they are free data. Scri is located at $\ttau=0$ and is umbilic in the conformal metric, just as the initial singularity was. We have done enough to show that the metric is at least $C^3$ there, and inductively we could get smoothness, but $S$ is not necessarily even about Scri.

 The fact that $S_\infty$ is free is equivalent to saying that the metric of Scri is free data but $S_3$ corresponds to part of the Weyl tensor (more accurately, to the time-derivative of the magnetic part of the Weyl tensor). See in this connection the discussion of vacuum solutions with data at Scri in \cite{fr}, and of the asymptotic behaviour of solutions in \cite{R2}:  the time-derivative of the magnetic part of the Weyl tensor is free data at Scri but not at the initial singularity. Of course, for the particular solutions considered here, which expand from initial isotropic singularities, the Weyl tensor at Scri is determined by the data at the bang, and it remains a possibility that $S_3$ is obliged to vanish at Scri. However, from the point of view of initial value problems, there is more free data at Scri than at the bang and if one evolves back from Scri with arbitrary choices of the data there then one cannot expect to arrive at an isotropic singularity at the beginning. 

Another way to see this last fact is to write the Einstein equations as a first-order system about Scri by introducing $R=\ttau^{-1}\tU$ and $S=e^\Sigma$  in place of (\ref{ks7}). The resulting first-order system is Fuchsian but does not satisfy the conditions for Theorem 1 of \cite{RS} (since it has to allow extra data at $O(\ttau^3)$ in $\Sigma$). This is one explanation for the absence of the inversion symmetry found for FRW.

\subsection{Bianchi class A models}
With fluid flow orthogonal to the surfaces of homogeneity, there is no loss of generality in assuming that the metric is diagonal in the invariant basis. Thus we may write the space-time metric as
\be
\tg=dt^2-R^2(e^{2\alpha}\sigma_1^{\;2}+e^{2\beta}\sigma_2^{\;2}+e^{2\gamma}\sigma_3^{\;2}),
\label{b1}
\ee
where the $\sigma_i$ are a basis of left-invariant one-forms for the isometry group. For Bianchi class A the one-forms can be assumed to satisfy the system
\bean
d\sigma_1&=&n_1\sigma_2\wedge\sigma_3\\
d\sigma_2&=&n_2\sigma_3\wedge\sigma_1\\
d\sigma_3&=&n_3\sigma_1\wedge\sigma_2
\eean
where the $n_i$ are chosen from $(1, -1, 0)$ (and inequivalent possibilities lead to different Bianchi types, see e.g. \cite{hom}). The parametrisation of the metric is chosen so that $R^6$ is the determinant of the physical spatial metric, so that
\[\alpha+\beta+\gamma=0.\]
We can, as in the previous two cases, solve the conservation equation to find
\[\rho R^4=m={\mathrm{constant}}\]
and we may introduce conformal time $\tau$ as in (\ref{R4}). 
Then the conformally-rescaled metric
\[g:=R^{-2}\tg=d\tau^2-(e^{2\alpha}\sigma_1^{\;2}+e^{2\beta}\sigma_2^{\;2}+e^{2\gamma}\sigma_3^{\;2})\]
is smooth wherever $(\alpha,\beta,\gamma)$ are. It gives the metric of the bang where $R=0$ and the metric of Scri where $R$ diverges.

It will be convenient to write
\[N_1=n_1e^{2\alpha},\;\;N_2=n_2e^{2\beta},\;\;N_3=n_3e^{2\gamma},\]
when we may write the Einstein equations at once in terms of $\tau$ as
\be
\alpha''+2\frac{R'}{R}\alpha'+A=0,
\label{b2}
\ee
and two cyclic permutations of this for $\beta$ and $\gamma$. Here
\[A=\frac{1}{3}(2N_1^{\;2}-N_2^{\;2}-N_3^{\;2}+2N_2N_3-N_1N_2-N_3N_1),\]
and we write $B$ and $C$ for the two cyclic permutations of this. Together with (\ref{b2}) we need the Hamiltonian constraint which is
\be
3\left(\frac{R'}{R}\right)^2=\frac{1}{2}((\alpha')^2+(\beta')^2+(\gamma')^2)+\frac{m}{R^2}+\Lambda R^2-F
\label{b3}
\ee
where
\be
F=-\frac{1}{4}(N_1^{\;2}+N_2^{\;2}+N_3^{\;2}-2N_2N_3-2N_3N_1-2N_1N_2),
\label{b4}
\ee
which is proportional to the Ricci scalar of the spatial metric. (Again the momentum constraint is vacuously satisfied.) By differentiating (\ref{b3}) we can obtain a second-order equation for $R$ which can be written as
\be
\frac{R''}{R}=-\frac{1}{6}((\alpha')^2+(\beta')^2+(\gamma')^2)-\frac{1}{3}F+\frac{2}{3}\Lambda R^2.
\label{b5}
\ee
Our first aim is to show that solutions exist with isotropic singularities. As in the previous subsection, this is accomplished by putting the equations into Fuchsian form. We introduce $U$ by $R=\tau e^U$ as in (\ref{ks7}) and then put the equations into first-order form by introducing variables $(X,Y,Z,W)$ according to
\bean
\alpha'&=&X\\
\beta'&=&Y\\
\gamma'&=&Z\\
U'&=&W
\eean
Then (\ref{b2}) and its permutations become
\bean
X'&=&-\frac{2}{\tau}X-2WX-A\\
Y'&=&-\frac{2}{\tau}Y-2WY-B\\
Z'&=&-\frac{2}{\tau}Z-2WZ-C\\
W'&=&-\frac{2}{\tau}W-W^2-\frac{1}{6}(X^2+Y^2+Z^2)-\frac{1}{3}F+\frac{2\Lambda}{3}\tau^2\;e^{2U}.
\eean
The combined system of eight equations is in Fuchsian form, and satisfies the assumptions of Theorem 1 of \cite{RS}. Thus solutions exist in a neighbourhood of $\tau=0$ with $(\alpha,\beta,\gamma)$ and $U$ freely specifiable there and $(X,Y,Z,W)$ all zero there.

The Hamiltonian constraint (\ref{b3}) can be written as
\[m=e^{2U}\left(3+6\tau W+\tau^2(3W^2-\frac{1}{2}(X^2+Y^2+Z^2)+F-\Lambda\tau^2\;e^{2U})\right),\]
when it can be read as relating the (positive) constant of integration $m$ to the initial value of $U$.

To study the behaviour at late times, we restrict to the Bianchi class A metrics other than type-IX. Recall that type-IX has $(n_1,n_2,n_3)=(1,1,1)$ and $F$ in (\ref{b4}) can have either sign. For all the other types, $F$ is non-positive. Thus from (\ref{b5}) $R''$ is positive, therefore so is $R'$ and solutions expand until $R$ diverges. There is a finite conformal-time blow up in $R$, at $\tau_F$ say, since (\ref{b3}) gives
\[3(R')^2\geq m+\Lambda R^4\]
and we can argue just as in subsection 3.2. Again $R$ will have a pole in $\tau$:
$R\sim (H(\tau_F-\tau))^{-1}$. Analogously to (\ref{ks8}), we introduce
\[Q=X^2+Y^2+Z^2-2F\]
then 
\be
Q'=-4\frac{R'}{R}(X^2+Y^2+Z^2),
\label{b6}
\ee
 so that $Q$ decreases and, for all time, 
\[0\leq Q\leq Q(0)=-6F(0).\]
Thus $X$, $Y$ and $Z$ are bounded for all time, so $\alpha$, $\beta$ and $\gamma$ grow at worse linearly in $\tau$ and, since there is only a finite amount of $\tau$-time, are also bounded and have limits at $\tau_F$. These finite limits define the metric of Scri. From (\ref{b2}) we obtain
\be
R^2X:=R^2\alpha'=-\int_0^{\tau}R^2(\sigma)A(\sigma)d\sigma,
\label{b7}
\ee
so that $\alpha'\rightarrow 0$ as $\tau\rightarrow\tau_F$. Thus $X$, $Y$ and $Z$ vanish at Scri and Scri is umbilic in the rescaled metric. For more differentiability, we mimic the calculation in the previous section. First set $\tR=R^{-1}$ so that (\ref{b3}) becomes
\[3(\tR')^2=\Lambda+Q\tR^2+m\tR^4,\]
and then with this and (\ref{b6}) and (\ref{b7}) we can obtain power series for $\alpha$, $\beta$, $\gamma$ and $R$. The rescaled metric will be smooth at Scri.

As in the previous subsection, $R$ is odd in $\tau$ and $(\alpha,\beta,\gamma)$ are even about the bang, but this doesn't necessarily hold at Scri, where the terms of order $\ttau^3$ in $(\alpha,\beta,\gamma)$ are not determined in the power series. Scri and the bang are both umbilic in the conformal metric, but are not equivalent, and in particular we don't have the symmetry interchanging them found for FRW. We do have solutions of the kind considered by Penrose \cite{P}, where the conformal metric is extendible through the bang and through Scri, but, at least with this matter model, it seems that we can't use Scri as a new bang, that is as the initial data surface for another cycle of expansion, without having the conformal metric not smooth, i.e. the normal-derivative of the magnetic Weyl tensor discontinuous.

\section*{Acknowledgements}
I am grateful to the Albert Einstein Institute, Golm, for hospitality while part of this work was carried out, and to Alan Rendall for useful conversations.

%
\end{document}